\documentclass[12pt]{article}%
\usepackage{amsmath,latexsym}
\usepackage{graphicx}
\usepackage{amsmath}
\usepackage{amsfonts}
\usepackage{amssymb}%
\begin{document}

\title{{Thin-shell wormholes from compact stellar objects}}
   \author{
Peter K. F. Kuhfittig*\\
\footnote{E-mail: kuhfitti@msoe.edu}
 \small Department of Mathematics, Milwaukee School of
Engineering,\\
\small Milwaukee, Wisconsin 53202-3109, USA}

\date{}
 \maketitle

\begin{abstract}\noindent
This paper introduces a new type of
thin-shell wormhole constructed from a
special class of compact stellar objects
rather than black holes.  The
construction and concomitant
investigation of the stability to
linearized radial perturbations commences
with an extended version of a regular
Hayward black hole.  Given the equation
of state $\mathcal{P}=\omega \sigma$,
$\omega <0$, for the exotic matter on
the thin shell, it is shown that whenever
the value of the Hayward parameter is
below its critical value, no stable
solutions can exist.  If the Hayward
parameter is allowed to exceed its
critical value, stable solutions can be
found for moderately sized thin shells.
Not only are the underlying structures
ordinary compact objects, rather than
black holes, the results are consistent
with the properties of neutron stars,
as well as other compact stellar
objects.  \\

\end{abstract}

\section{Introduction}\label{S:Introduction}

A highly effective method for describing or
mathematically constructing a class of
spherically symmetric wormholes using the now
standard cut-and-paste technique was proposed
by Visser in 1989 \cite{mV89}.  The construction
calls for grafting two black-hole spacetimes
together, resulting in a \emph{thin-shell
wormhole}.  By starting with a regular Hayward
black hole, it is proposed in this paper that
the construction can be extended to massive
compact objects.

Apart from a number of forerunners, the concept
of a traversable wormhole suitable for
interstellar travel was first proposed by
Morris and Thorne \cite{MT88}.  It turned out
that a wormhole could only be held open by
violating the null energy condition, defined as
follows:
\begin{equation}
   T_{\alpha\beta}k^{\alpha}k^{\beta}\ge 0
\end{equation}
for all null vectors $k^{\alpha}$.  Matter
that violates this condition came to be called
``exotic."  In particular, for the radial
outgoing null vector $(1,1,0,0)$, the violation
takes on the form $T_{\alpha\beta}k^{\alpha}k^{\beta}
=\rho +p<0$.  (Here $T_{00}=\rho$, the energy
density, and $T_{11}=p$, the radial pressure.)

We will assume in this paper that the exotic
matter on the shell satisfies a certain equation
of state (EoS).  In an earlier paper, Eiroa
\cite{eE09} assumed the generalized Chaplygin
EoS $\mathcal{P}=A/|\sigma|^{\alpha}$, where
$\sigma$ is the energy density of the shell
and $\mathcal{P}$ is the surface pressure.
Kuhfittig \cite{pK12} investigated the possible
stability of thin-shell wormholes constructed
from several spacetimes using the simpler EoS
$\mathcal{P}=\omega\sigma$, $\omega <0$.  Since
it resembles the EoS of a perfect fluid, this
choice seems more natural and will therefore be
employed in this paper.

As will be seen below, the energy density
$\sigma$ of the thin shell is negative.
Moreover, given that the shell is assumed to
be infinitely thin, the radial pressure $p$ is
zero.  So $\rho +p=\sigma +0<0$, so that the
null energy condition is automatically
violated.

Our final goal in this paper is to determine
criteria for making this new type of wormhole
stable to linearized radial perturbations.

\section{Thin-shell wormhole construction}
   \label{S:construction}\noindent
Consider the line element
\begin{equation}\label{E:line1}
ds^2 = -f(r) dt^2 + [f(r)]^{-1}dr^2
  + h(r)(d\theta^2+\sin^2\theta d\phi^2),
\end{equation}
where $f(r)$ is a positive function of $r$.  As in
Ref. \cite{PV95}, the construction begins with two
copies of a black-hole spacetime and removing from
each the four-dimensional region
\begin{equation}\label{E:remove}
  \Omega^\pm = \{r\leq a\,|\,a>r_h\},
\end{equation}
where $r=r_h$ is the (outer) event horizon of the black
hole.  Now identify (in the sense of topology) the
time-like hypersurfaces
\begin{equation}
  \partial\Omega^\pm =\{r=a\,|\,a>r_h\}.
\end{equation}
The resulting manifold is geodesically complete and
possesses two asymptotically flat regions connected
by a throat.  Next, we use the Lanczos equations
\cite{mV89, eE09, pK12, PV95, LC04, ER04, TSE06, RKC06,
RKC07, RS07, LL08}
\begin{equation}\label{E:Lanczos}
  S^i_{\phantom{i}j}=-\frac{1}{8\pi}\left([K^i_{\phantom{i}j}]
   -\delta^i_{\phantom{i}j}[K]\right),
\end{equation}
where $S^i_{\phantom{i}j}$ is the surface stress-energy
tensor, $K^i_{\phantom{i}j}$ is the extrinsic curvature
tensor, $[K_{ij}]=K^{+}_{ij}-K^{-}_{ij}$ and $[K]$ is the
trace of $K^i_{\phantom{i}j}$.  In terms of the surface
energy density $\sigma$ and the surface pressure
$\mathcal{P}$, $S^i_{\phantom{i}j}=\text{diag}(-\sigma,
\mathcal{P}, \mathcal{P})$.  The Lanczos equations now yield
\begin{equation}
  \sigma=-\frac{1}{4\pi}[K^\theta_{\phantom{\theta}\theta}]
\end{equation}
and
\begin{equation}\label{E:LanczosP}
  \mathcal{P}=\frac{1}{8\pi}\left([K^\tau_{\phantom{\tau}\tau}]
    +[K^\theta_{\phantom{\theta}\theta}]\right).
\end{equation}

A dynamic analysis can be obtained by letting the radius $r=a$
be a function of time \cite{PV95}.  As a result,
\begin{equation}\label{E:sigma}
\sigma = - \frac{1}{2\pi a}\sqrt{f(a) + \dot{a}^2}
\end{equation}
and
\begin{equation}\label{E:P}
  \mathcal{P} =  -\frac{1}{2}\sigma + \frac{1}{8\pi
}\frac{2\ddot{a} + f^\prime(a) }{\sqrt{f(a) + \dot{a}^2}}.
\end{equation}
Here the overdots denote the derivatives with respect to
proper time $\tau$.

It is easy to check that $\sigma$ and $\mathcal{P}$ obey
the conservation equation
\begin{equation}
   \frac{d}{d\tau}(\sigma a^2)
       +\mathcal{P}\frac{d}{d\tau}(a^2)=0,
\end{equation}
which can also be written in the form
\begin{equation}\label{E:conservation}
  \frac{d\sigma}{da} + \frac{2}{a}(\sigma+\mathcal{P}) = 0.
               \end{equation}

To perform a stability analysis, we must first note that
for a static configuration of radius $a$, we have
$\dot{a}=0$ and $\ddot{a}=0$.  We must also consider
linearized fluctuations around a static solution
characterized by the constants $a_0$, $\sigma_0$, and
$\mathcal{P}_0$.  Now, given the EoS
$\mathcal{P}=\omega\sigma$, Eq. (\ref{E:conservation})
can be solved by separation of variables to yield
\begin{equation*}
  |\sigma(a)|=|\sigma_0|\left(\frac{a_0}{a}
    \right)^{2(\omega+1)},
\end{equation*}
where $\sigma_0=\sigma(a_0)$.  The solution can
therefore be written as
\begin{equation}\label{E:sigmaexplicit}
  \sigma(a)=\sigma_0\left(\frac{a_0}{a}
  \right)^{2(\omega+1)},\quad \sigma_0=\sigma(a_0).
\end{equation}

The next step is to rearrange Eq. (\ref{E:sigma})
to obtain the ``equation of motion"
\begin{equation}
\dot{a}^2 + V(a)= 0,
\end{equation}
where $V(a)$ is the potential defined  as
\begin{equation}\label{E:Vdefined}
V(a) =  f(a) - \left[2\pi a \sigma(a)\right]^2.
\end{equation}
Taylor-expanding $V(a)$ around $a=a_0$, we obtain
\begin{eqnarray}
V(a) &=&  V(a_0) + V^\prime(a_0) ( a - a_0) +
\frac{1}{2} V^{\prime\prime}(a_0) ( a - a_0)^2
+\text{higher-order terms}.
\end{eqnarray}
Since we are linearizing around $a=a_0$, we require that
$V(a_0)=0$ and $V'(a_0)=0$.  Since the higher-order
terms are considered negligible, the configuration
is in stable equilibrium if $V''(a_0)>0$.

\section{The regular Hayward black hole}

The following line element describes a spherically
symmetric regular (nonsingular) black hole:
\begin{equation}\label{E:line2}
ds^{2}=-\left(1-\frac{2mr^2}{r^3+2ml^2}\right)dt^{2}
   +\left(1-\frac{2mr^2}{r^3+2ml^2}\right)^{-1}dr^{2}
+r^{2}(d\theta^{2}+\text{sin}^{2}\theta\,
d\phi^{2}),
\end{equation}
Introduced by Hayward \cite{sH06}, this is referred to
as a \emph{Hayward black hole} in Refs. \cite{HOH, SM}.
Line element (\ref{E:line2}) contains two free
parameters, $l$ and $m$; $l$ is called the \emph
{Hayward parameter}, while $m$ will be interpreted as
the mass of the black hole.  Thin-shell wormholes from
Hayward black holes are discussed in Refs. \cite{HOH, SM}.

It is apparent that for large $r$, the Hayward black
hole becomes a Schwarzschild spacetime.
To study the behavior for small $r$, we first rewrite
line element (\ref{E:line2}) in a form that is
particularly convenient for later analysis:
\begin{multline}\label{E:line3}
ds^{2}=-\left(1-\frac{2(r/m)^2}{(r/m)^3+2(l/m)^2}\right)dt^{2}
   +\left(1-\frac{2(r/m)^2}{(r/m)^3+2(l/m)^2}\right)^{-1}dr^{2}\\
+r^{2}(d\theta^{2}+\text{sin}^{2}\theta\,
d\phi^{2}).
\end{multline}
 Writing the $g_{tt}$-term in the form
 \[
   1-\frac{(r/m)^2}{(l/m)^2}+\left(\frac{r/m}{l/m}\right)^5
   \frac{\frac{1}{2}\frac{l}{m}}
       {1+\frac{1}{2}\frac{(r/m)^3}{(l/m)^2}},
 \]
it follows that for small $r$,
\begin{equation}\label{E:order}
  f(r)=1-\frac{(r/m)^2}{(l/m)^2}
  +\mathcal{O}\left(\frac{r/m}{l/m}\right)^5.
\end{equation}
The point is that for small $r$, the Hayward black
hole has the form of a de Sitter black hole,
immediately raising the question whether the
resulting thin-shell wormhole could ever be stable
to linearized radial perturbations.  Ref.
\cite{pK12} discusses the de Sitter case $f(r)=
1-2m/r-\frac{1}{3}\Lambda r^2$, where $\Lambda >0$
is the cosmological constant.  A stable solution
requires that $\frac{1}{3}\Lambda m^2<\frac{1}{27}$.
Now, Refs. \cite{HOH, SM} note that for a Hayward
black hole, there is a critical value $(l/m)^2=
16/27$ corresponding to a regular extremal black
hole.  The smaller value $(l/m)^2<16/27$ admits a
double horizon.  If $(l/m)^2>16/27$, there is no
event horizon at all, a point that will be
addressed later.  For now, it is sufficient to
observe that the values for $1/(l/m)^2$ in Eq.
(\ref{E:order}) are much larger than the allowed
value of $1/27$ from Ref. \cite{pK12}.  So
assuming the EoS $\mathcal{P}=\omega\sigma$, there
are no stable solutions for the de Sitter case.
(Since the Hayward black hole becomes a
Schwarzschild spacetime, nor are there stable
solutions for large $r$ \cite{pK12}.)

\section{Extending the Hayward black hole}

At this point we need to return to Sec.
\ref{S:construction}.  In particular, making use of
Eqs. (\ref{E:sigma}), (\ref{E:sigmaexplicit}), and
(\ref{E:Vdefined}), we obtain
\begin{equation}\label{E:potential}
   V(a)=1-\frac{2(a/m)^2}{(a/m)^3+2(l/m)^2}
   -\left[1-\frac{2(a_0/m)^2}{(a_0/m)^3+2(l/m)^2}
   \right]\left(\frac{a_0}{a}\right)^{2+4\omega}.
\end{equation}
Evidently, $V(a_0)=0$.  To meet the condition
$V'(a_0)=0$, we differentiate $V(a)$ in Eq.
(\ref{E:potential}) and solve for $\omega$:
\begin{equation}\label{E:omega}
   \omega =-\frac{1}{2}+\frac{1}{4}
   \frac{[(a_0/m)^3+2(l/m)^2](4)(a_0/m)^2-6(a_0/m)^5}
   {[(a_0/m)^3+2(l/m)^2][(a_0/m)^3+2(l/m)^2-2(a_0/m)^2]}.
\end{equation}

Since our spacetime is approximately de Sitter
only for small $r$, we will confine ourselves to
relatively small values of the shell radius.  As
already noted, values of the Hayward parameter
below the critical value yield a black hole with
two event horizons, but the resulting thin-shell
wormholes are unstable.  As we will see in the
next section, however, the calculations using
the Hayward line element do not require any
particular restriction on $l/m$.  In fact,
stable solutions can be obtained for values
above the critical value.  The implication is
that even though the line element retains its
original form, we are no longer dealing with
a black hole but instead with an ordinary,
possibly compact, stellar object.  More
precisely, line element (\ref{E:line2}) becomes
\begin{equation}\label{E:line4}
ds^{2}=-\left(1-\frac{2M(r)}{r}\right)dt^{2}
   +\left(1-\frac{2M(r)}{r}\right)^{-1}dr^{2}
+r^{2}(d\theta^{2}+\text{sin}^{2}\theta\,
d\phi^{2}),
\end{equation}
where
\begin{equation}
    M(r)=\frac{r^3m}{r^3+2ml^2}, \quad
    M(0)=0,
\end{equation}
showing that Eq. (\ref{E:line4}) represents an
ordinary mass such as a particle or a star
\cite{MTW}.  (We will see below that we are
dealing with a special class of compact stellar
objects.)

We are going to obtain some stable solutions in
the next section.

\section{Stable solutions; compact stellar objects}

To illustrate the type calculation required,
suppose we choose $a_0/m=2$ and $l/m=
\sqrt{16/27}$, the critical value.  Then from
Eq. (\ref{E:omega}), we get $\omega=-1.534274$
and $2+\omega =-4.13710$.  Eq. (\ref{E:potential})
then gives
\begin{equation}
   V(a)=1-\frac{2(a/m)^2}{(a/m)^3+2(16/27)}-
   \left[\frac{2(2^2)}{2^3+2(16/27)}\right]
   \left(\frac{2}{a/m}\right)^{-4.13710}.
\end{equation}
 The resulting graph in Fig. 1 is concave
\begin{figure}[tbp]
\begin{center}
\includegraphics[width=0.8\textwidth]{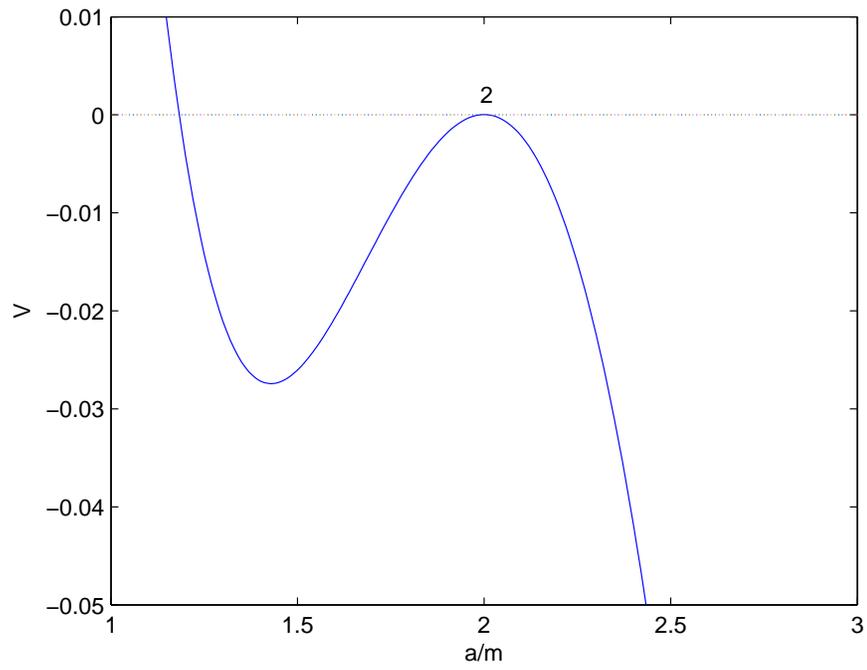}
\end{center}
\caption{The wormhole is unstable.}
\end{figure}
down around $a_0/m=2$, showing that the wormhole is
unstable.

As another example, if $a_0/m=3.5$ and $l/m=1.7$,
we obtain $2(l/m)^2=5.78$, $2+4\omega =-0.6528$,
and
\begin{equation}
   V(a)=1-\frac{2(a/m)^2}{(a/m)^3+5.78}-
   \left[1-\frac{2(3.5)^2}{3.5^3+5.78}\right]
   \left(\frac{3.5}{a/m}\right)^{-0.6528}.
\end{equation}
The graph, shown in Fig. 2, is concave up
\begin{figure}[tbp]
\begin{center}
\includegraphics[width=0.8\textwidth]{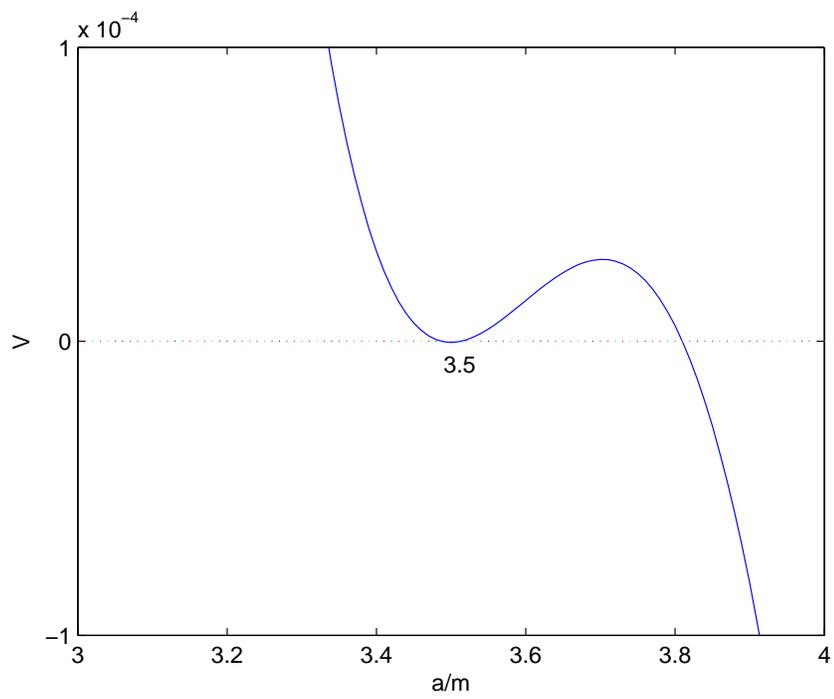}
\end{center}
\caption{The wormhole is stable.}
\end{figure}
around $a_0/m=3.5$.  So the wormhole is stable.

Table 1 provides an overview of the results for
various values of $a_0/m$, starting with
$a_0/m=1.5$ by showing the dependence on the
Hayward parameter: to ensure stability, a
larger value of $a_0/m$ requires a larger
value of $l/m$.  (For $a_0/m\le 1$, we get
only unstable solutions.)
\begin{table}
  \begin{center}
    \begin{tabular}{|c|c|c|}
    \hline
    $\boldsymbol{a_0/m}$ & $\boldsymbol{l/m}$ &
       \textbf{Status}  \\ \hline \hline
    1.5 & $\sqrt{16/27}$ & unstable \\ \hline
    1.5 & 0.8 & stable \\ \hline \hline
    2 & 0.9 & unstable \\ \hline
    2 & 1.0 & stable \\ \hline \hline
    2.5 & 1.1 & unstable \\ \hline
    2.5 & 1.2 & stable \\ \hline \hline
    3 & 1.3 & unstable \\ \hline
    3 & 1.4 & stable \\ \hline \hline
    3.5 & 1.6 & unstable \\ \hline
    3.5 & 1.7 & stable \\ \hline \hline
    4 & 1.9 & unstable \\ \hline
    4 & 2.0 & stable \\ \hline \hline
    4.5 & 2.2 & unstable \\ \hline
    4.5 & 2.3 & stable \\ \hline \hline
    5 & 2.6 & unstable \\ \hline
    5 & 2.7 & stable \\ \hline
    \end{tabular}
    \caption{Larger values of $a_0/m$ require
    larger values of $l/m$ for stability.}
  \end{center}
\end{table}

Since we are dealing with compact objects rather
than black holes, we need to check the physical
plausibility.  Consider a typical neutron star
having a radius ranging from 11 km to 11.5 km
and a mass of $2M_{\odot}\approx 3\, \text{km}$.
Then $a_0/m$ ranges from 3.67 to 3.83, which
are well within the values listed in Table 1.
In fact, it is theoretically possible to have
a stable thin shell right above the surface
with $a_0/m=4$ and $l/m=2.0$ according to
Table 1.

Other compact objects such as quark stars or
strange stars could allow even smaller values
for $a_0/m$.

\section{Conclusion}
This paper introduces a new type of thin-shell
wormhole constructed from a special class of compact
stellar objects rather than black holes.  This
finding follows from a discussion of the stability
to linearized radial perturbations of an extended
version of a regular Hayward black hole.  Assuming
the equation of state $\mathcal{P}=\omega \sigma$,
$\omega <0$, for the exotic matter on the thin
shell, it is shown that whenever the value of the
Hayward parameter is below its critical value, no
stable solutions can exist.  Stable solutions are
obtained, however, if $l/m$ is allowed to exceed
the critical value, thereby eliminating the event
horizon of the black hole.  The resulting
underlying structure supporting the thin-shell
wormhole is a compact object rather than a
black hole.  The findings are consistent with the
properties of neutron stars, as well as other compact
stellar objects.


\begin{thebibliography}{20}

\bibitem{mV89}M. Visser, ``Traversable wormholes from
   surgically modified Schwarzchild spacetimes,"
   Nucl. Phys. B {\bf 328}, 203 (1989).
\bibitem{MT88}M.S. Morris and K.S. Thorne, ``Wormholes
   in spacetime and their use for interstellar travel:
   A tool for teaching general relativity,"  Amer. J. Phys.
   \textbf{56}, 395 (1988).
\bibitem{eE09}E.F. Eiroa, ``Thin-shell wormholes with
   a generalized Chaplygin gas," Phys. Rev. D {\bf 80},
   044033 (2009).
\bibitem{pK12}P.K.F. Kuhfittig, ``The stability of
   thin-shell wormholes with a phantom-like equation of
   state," Acta Phys. Polon. B {\bf 41}, 2017 (2010).
\bibitem{PV95}E. Poisson and M. Visser, ``Thin-shell
   wormholes: Linearized stability," Phys. Rev. D
  {\bf 52}, 7318 (1996).
\bibitem{LC04}F.S.N. Lobo and P. Crawford, ``Linearized
   stability analysis of thin-shell wormholes with a
   cosmological constant," Class. Quant. Grav. {\bf 21},
   391 (2004).
\bibitem{ER04}E.F. Eiroa and G.E. Romero, ``Linearized
stability of charged thin-shell wormholes," Gen. Rel.
   Grav. {\bf 36}, 651 (2004).
\bibitem{TSE06}M. Thibeault, C. Simeone, and E.F. Eiroa,
   ``Thin-shell wormholes in Einstein-Maxwell theory
   with a Gauss-Bonnet term," Gen. Rel. Grav.
   {\bf 38}, 1593 (2006).
\bibitem{RKC06}F. Rahaman, M. Kalam, and S. Chakraborty,
   ``Thin shell wormholes in higher dimensional
   Einstein-Maxwell theory," Gen. Rel. Grav.
   {\bf 38}, 1687 (2006).
\bibitem{RKC07}F. Rahaman, M.  Kalam, and S. Chakraborty,
   ``Gravitational lensing by a stable C-field
   wormhole," Chin. J. Phys. {\bf 45}, 518 (2007).
\bibitem{RS07}M.G. Richarte and C. Simeone,
   ``Thin-shell wormholes supported by ordinary
   matter in Einstein-Gauss-Bonnet gravity,"  Phys.
   Rev. D {\bf76}, 087502 (2007).
\bibitem{LL08}J.P.S. Lemos and F.S.N. Lobo, ``Plane
   symmetric thin-shell wormholes: Solutions and
   stability," Phys. Rev D {\bf 78}, 044030 (2008).
\bibitem{sH06}S.A. Hayward, ``Formation and evaporation
   of nonsingular black holes," Phys. Rev. Lett.
   {\bf 96}, 031103 (2006).
\bibitem{HOH}M. Halilsoy, A. Ovgun, and S. Habib Mazharimousavi,
   ``Thin-shell wormholes from regular Hayward black hole,"
   Eur. Phys. J. C {\bf 74}, 2796 (2014).
\bibitem{SM}M. Sharif and S. Mumtaz, ``Stabilty of
regular Hayward thin-shell wormholes," Ad. High Energy Phys.
   \textbf{2016}, 2868750, (2016).
\bibitem{MTW}C.W. Misner, K.S.  Thorne, and J.A. Wheeler,
   Gravitation. W. Freeman and Company, New York (1973)
   p. 608.

 \end{thebibliography}
\end{document}